\documentclass[twocolumn]{aastex62}
\usepackage{color}
\received{}
\revised{}
\accepted{}
\submitjournal{}

\shorttitle{Near and far ultraviolet properties of the M31 Bulge}
\shortauthors{Leahy et al.}

\begin{document}

\title{ASTROSAT/UVIT Near and Far Ultraviolet Properties of the M31 Bulge}

\correspondingauthor{Denis Leahy}
\email{leahy@ucalgary.ca}

\author{Denis Leahy}
\affiliation{Department of Physics and Astronomy, University of Calgary, Calgary, AB T2N 1N4, Canada}

\author{Cole Morgan}
\affiliation{Department of Physics and Astronomy, University of Calgary, Calgary, AB T2N 1N4, Canada}

\author{Joseph Postma}
\affiliation{Department of Physics and Astronomy, University of Calgary, Calgary, AB T2N 1N4, Canada}

\author{Megan Buick}
\affiliation{Department of Physics and Astronomy, University of Calgary, Calgary, AB T2N 1N4, Canada}


\begin{abstract}
 AstroSat has surveyed M31 with the UVIT telescope during 2017 to 2019. 
 The central bulge of M31 was observed in 
2750-2850 A, 2000-2400 A, 1600-1850 A, 
 1450-1750 A, and 1200-1800 A filters. 
 A radial profile analysis, averaged along elliptical contours which approximate the bulge shape, was carried out in 
 each filter. The profiles follow a Sersic function with an excess for the inner $\sim8\arcsec$ 
 in all filters, or can be fit with two Sersic functions (including the excess). 
 The ultraviolet colours of the bulge are found to change systematically with 
 radius, with the center of the bulge bluer (hotter). 
 We fit the UVIT spectral energy distributions (SEDs) for the whole bulge and for 10 elliptical annuli with 
 single stellar population (SSP) models.  
A combination of two SSPs fits the UVIT SEDs much better than one SSP, and three SSPs fits the data
best. The properties of the three SSPs are age, metallicity ($Z$) and mass of each SSP. 
The best fit model is a dominant old, metal-poor ($10^{10}$yr, log($Z/Z_{\odot})=-2$,  with 
$Z_{\odot}$ the solar metallicity)  population plus a 15\% 
contribution from an intermediate ($10^{9.5}$ yr, log($Z/Z_{\odot})=-2$) population plus a small contribution 
($\sim$2\%) from a young high-metallicity ($10^{8.5}$ yr, log($Z/Z_{\odot})=-0.5$) population. 
The results are consistent with previous studies of M31 in optical: both reveal an active merger history for M31.
\end{abstract}



\section{Introduction} \label{sec:intro}

The spiral galaxy Andromeda, also known as M31, is the nearest such galaxy to our own. 
Our external vantage point makes it feasible to study aspects of M31 that are difficult to study in our own galaxy. 
Interstellar extinction is not as major a factor when studying M31 as it is studying the Milky Way, owing to the 
former's inclination that allows observation of the whole disk of the galaxy. Studies of large numbers of stars 
have been done in great detail for the Milky Way, but distances often possess high uncertainty, due in part  to extinction. 
An advantage of studying objects in M31 is that it is at a well known distance (785$\pm25$ kpc, \citealt{2005McConnachie}). 
The intrinsic brightness of many objects can therefore be more precisely measured than galactic sources.
M31 has been observed in optical on numerous occasions. The highest resolution observations were carried out with 
the Hubble Space Telescope (HST), including the Pan-chromatic Hubble Andromeda Treasury (PHAT) survey 
\citep{PHAT}.
In near and far ultraviolet (NUV and FUV), the GALEX instrument \citep{2005ApJ...619L...1M} has surveyed M31.

The AstroSat mission, launched on September 28th, 2015,  has carried out a survey of M31 in near and far UV. 
AstroSat is an orbiting observatory equipped with five instruments:  the UV Imaging 
Telescope (UVIT) for visible and UltraViolet (UV); the Soft X-ray Telescope (SXT), Large Area Proportional Counters 
(LAXPC) and Cadmium-Zinc-Telluride Imager (CZTI) instruments for soft through hard X-rays; and 
the Scanning Sky Monitor (SSM), an X-ray survey instrument \citep{singh+2014}. 

\begin{figure*}[htbp]
    \centering
    \epsscale{0.9}
 \plotone{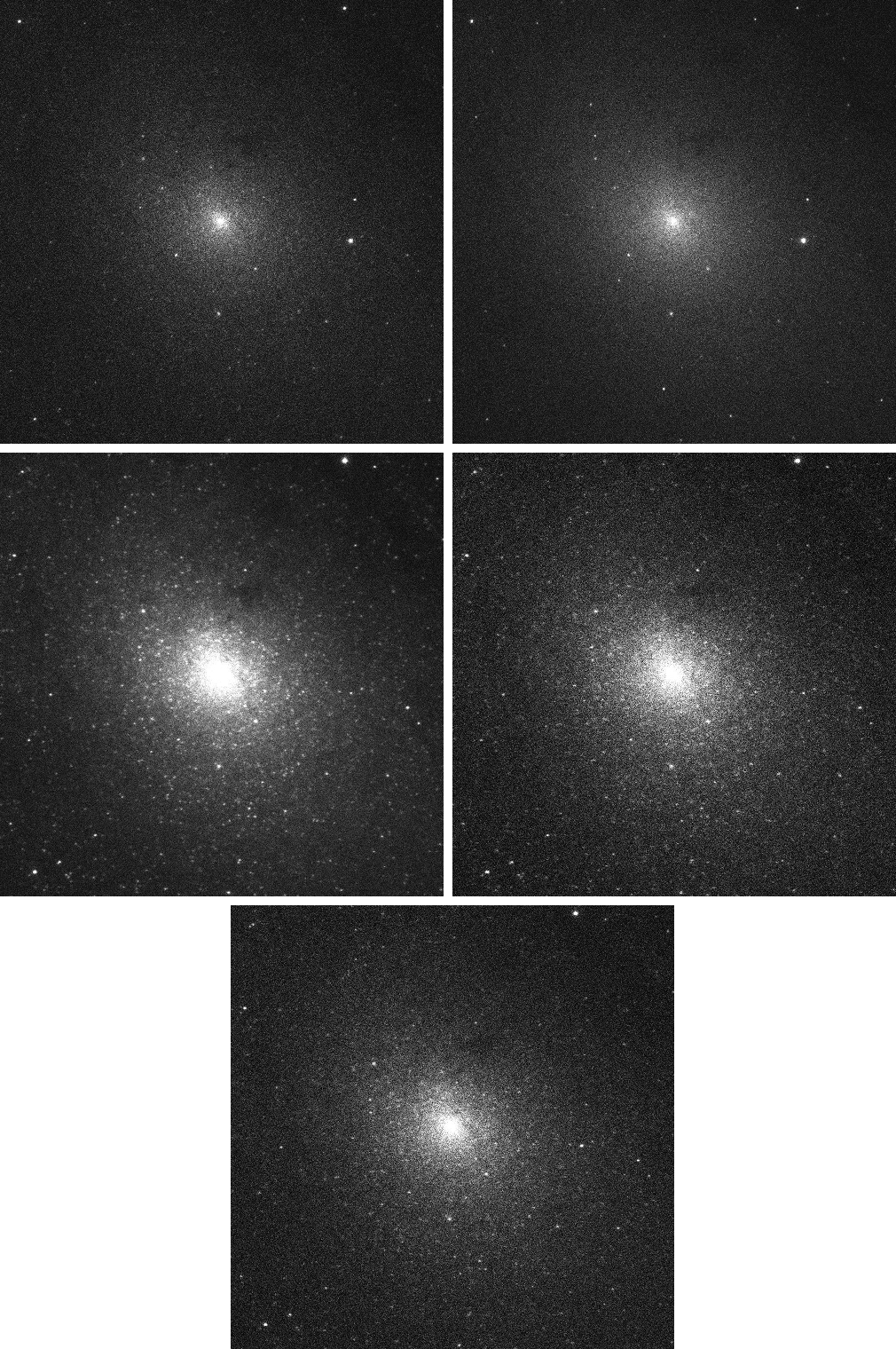}
     \figurenum{1}
   \label{fig:bulgeimages}
  \caption{The five processed images used in this analysis. Top Left: N279N\_A, top right: N219M\_A, middle left: F148W\_M, middle right: F169M\_B, bottom: F172M\_M. The \_A, \_B and \_M refer to epoch A, epoch B and merged images,
  respectively. The images are 410" by 410" across, and centred on 00h 42m 44.34s ra., +41d 16' 08.44" dec.}
\end{figure*}

Analysis of the M31 UVIT survey observations have been presented in part by \citet{2018AJ....156..269L},  
\citet{2020ApJS..247...47L},  \citet{2020ApJS..250...23L}, \citet{2020arXiv201202727L} and 
Leahy et al. 2021 (submitted for publication). 
These papers have dealt with resolved stellar objects, including analysis of UV brightest stars in the bulge, 
the M31 UVIT point source catalog, matching UVIT point sources with Chandra sources in M31, 
improvements in astrometry and photmetry for the M31 survey, first results from matching UVIT sources with HST/PHAT 
sources in the NE spiral arms of M31, and a study of FUV variable sources in M31 using a new second epoch 
observation of the central field of M31.
The UVIT observations for the M31 survey are described in the M31 UVIT point source catalog paper 
\citep{2020ApJS..247...47L}. 
19 UVIT fields of 28 arcmin diameter were required to cover the M31 area (Fig. 2 of \citealt{2020ApJS..247...47L}).

The current analysis is primarily concerned with the unresolved ultraviolet sources in M31's central bulge.
The bulge of a galaxy is the central concentrated set of stars which is ellipsoidal in shape, 
smoothly distributed and consists of a substantial fraction of the galaxy's stars. 
UVIT's multiple filters within the NUV and FUV bands allow for an in-depth look at the UV colours of the bulge. 
In Sections \ref{sec:obs} and \ref{sec:anal}, the observations and data analysis methods are described. 
The results are reviewed in Section \ref{sec:results}  including radial profile analysis and UV colour change 
of the bulge with radius. 
In Section \ref{sec:disc}  synthetic colour magnitude diagrams (CMDs) are used to model the colour change of the bulge 
with radius and constrain the stellar populations in the bulge. 
We close with a brief summary.

\section{Observations}  \label{sec:obs}

The UVIT instrument onboard AstroSat has high spatial resolution ($\simeq$1 arcsec), a 28 arcminute field of view, 
and was capable of observing in a variety of FUV and NUV filters \citep{2017AJ....154..128T}. 
However, partway through observations of M31, the NUV detector of UVIT failed, resulting in partial coverage of M31 
in NUV. The entire galaxy has been observed in the FUV, and the central bulge is included in the portion with NUV data.
The central bulge of M31 is entirely contained within Field 1 of the UVIT Andromeda survey. 
The bulge of M31 was observed by UVIT in 2017 \citep{2018AJ....156..269L} and at a second epoch in 2019.
New in-orbit calibrations of the UVIT instrument have been carried out by \citep{2020AJ....159..158T} 
and are applied here to the 2017 and 2019 observations.

\begin{figure}[htbp]
    \centering
    \epsscale{1.2}
 \plotone{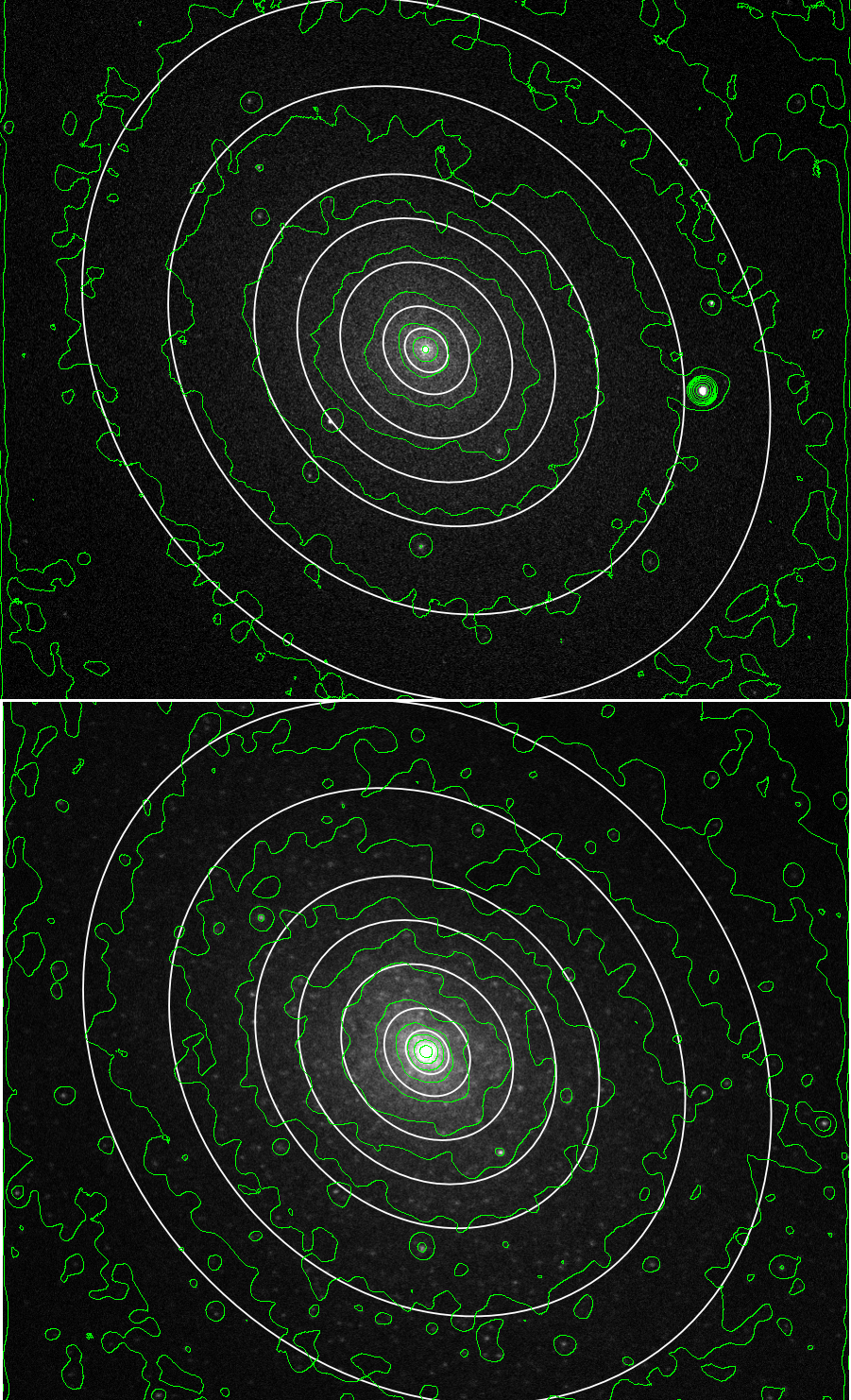}
    \figurenum{2}
    \label{fig:ellipsescontours}
\caption{The bulge of M31 in N279N (top) and F148W (bottom) with contours drawn in green using the ds9 image viewer, 
and a subset, drawn in white, selected from the 240 ellipses used to create the elliptical profiles. The images are 370$^{\prime\prime}$ E-W by 330$^{\prime\prime}$ N-S.}
\end{figure}

The basic data processing was carried out using CCDLab (Postma and Leahy, 2017), with updated astrometry calibration 
from \citet{2020PASP..132e4503P}, and updated photometry as described in  \citet{2020arXiv201202727L} and 
Leahy et al. 2021 (submitted for publication). 
M31 Field 1 (the bulge field) was observed by UVIT instrument in 5 FUV and NUV filters. 
From longest to shortest wavelength, these are: N279N, N219M, F172M, F169M, and F148W, where the label contains the
central wavelength (in nm) of each filter and band\footnote{N is for NUV, F is for FUV.  
The filter transmission curves are given in \citealt{2017AJ....154..128T}.}. 

\begin{figure*}[htbp]
    \centering
    \epsscale{1.1}
 \plottwo{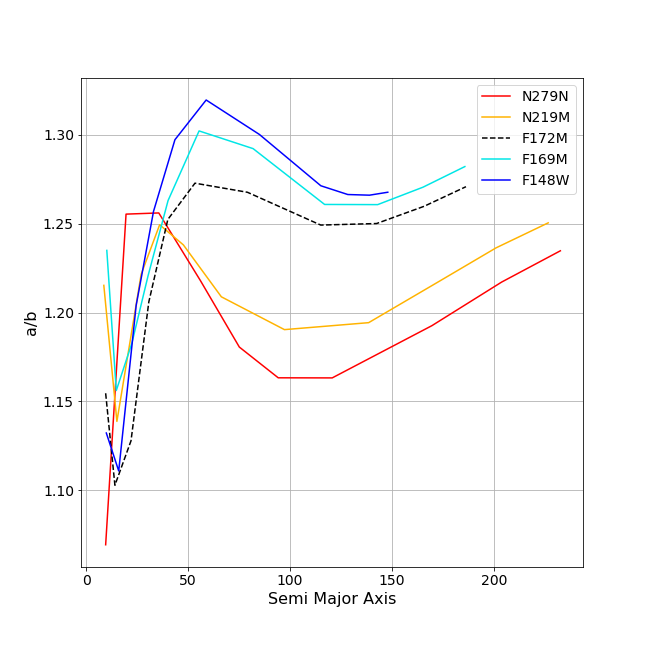}{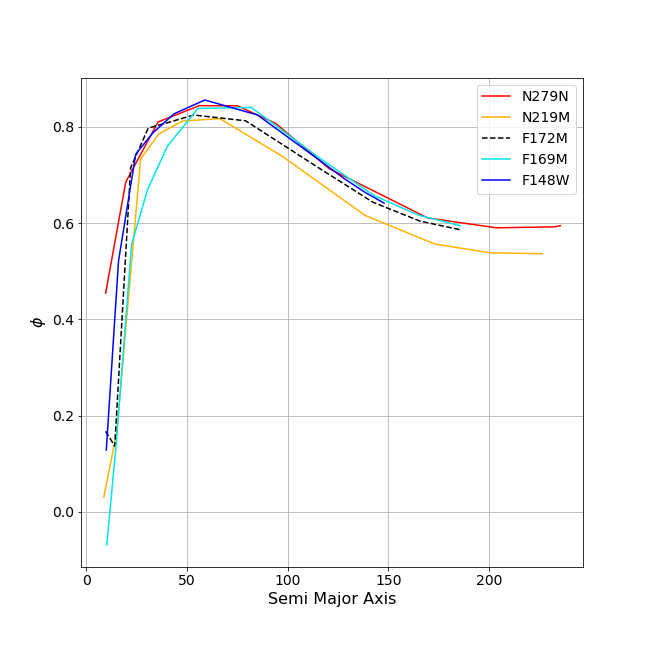}
    \figurenum{3}
\label{fig:GaussFitsAxes}
\caption{Left:  Ellipticity, (semi-major axis a)/(semi-minor axis b), of the bulge as a function of semi-major axis in pixel units, 
derived from Gaussian fits. Right: Inclination (radians, CW from W) of the bulge as a function of semi-major axis in
pixel units. 1 pixel is 0.4168$\arcsec$. }
\end{figure*}

The images were taken across two epochs, labelled A for the 2017 observation and B for the 2019 observation. 
The first set of observations included N279N, N219M, F172M, and F148W images, while the second set included 
F172M,  F169M and F148W images. 
Merged images, here labelled "M", were produced for filters F172M and F148W which had data from both A and B epochs.
The merged images were created by adding the counts images for A and B, adding the exposure maps for A and B,
then dividing the counts images by the exposure images to create a count/s image from which fluxes can be 
extracted using the UVIT calibrations (updated in \citealt{2020AJ....159..158T}).
  
The full 3-colour image of Field 1 made from the 2017 epoch N279N, N219M and F148W images is shown in 
\citet{2018AJ....156..269L}.
The image includes the bright central bulge, which is likely dominated by unresolved FUV and NUV emission from large numbers of low-luminosity stars in the bulge.
Field 1 contains other prominent features, including resolved and unresolved emission from numerous stars in the inner
spiral arms of M31, dust lanes which are visible as dark lanes, and several foreground stars, which can be identified 
using either Gaia parallax \citep{2020ApJS..250...23L} or colour-magnitude diagram analysis \citep{2020ApJS..247...47L}. 

The central part of Field 1 is the area that contains the central bulge of M31, here defined as the inner $\simeq$7 
arcminute diameter part of the 28 arcminute diameter full area of Field 1.
The area studied here is shown in Figure 1, 
with the images of the bulge shown for the five different NUV and FUV filters.

Due to the low inclination of M31, a few dust lanes are visible, e.g. NW of the bulge by $\sim$80 arcmin. 
The amount of area covered by the dust lanes is not large, so we did not correct for the reduction in flux caused 
by these dust lanes in our analysis. 
However, we did test an analysis which accounts for the bright point sources in the images. 
We masked the bright point sources and replaced them with the average values of surrounding pixels.
We compared the brightness profiles derived with and without the point sources and found there was no significant
difference. Thus we decided to leave the point sources in the images for the final analysis. 

\section{Data Analysis}  \label{sec:anal}

\subsection{Elliptical brightness profile extraction}

Figure~\ref{fig:ellipsescontours} illustrates the ellipticity of the central bulge of M31 at the longest NUV and
shortest FUV wavelengths. 
Shape changes of the bulge across different wavelengths are investigated here by fitting elliptical Gaussian 
functions to the images using CCDLab for a set of square bounding boxes of different sizes all centred on the bulge. 
The centre of the bulge was chosen as the brightest pixel in the core in the majority of filters. 
Figure~\ref{fig:GaussFitsAxes}  shows 
the ellipticity $(a/b)$ and inclination $(\phi)$  
as a function of major axis $(a)$. 
Inclination is measured in radians clockwise from the EW axis. 
Data from bounding boxes  $<$40 pixels half-width are not shown. 
This corresponds to a cutoff of 4.2$\arcsec$, the small central part of
the bulge, which has a complex structure (e.g. \citealt{2017PhDT.......170L}).

\begin{figure*}[htbp]
    \centering
    \epsscale{1.15}
 \plottwo{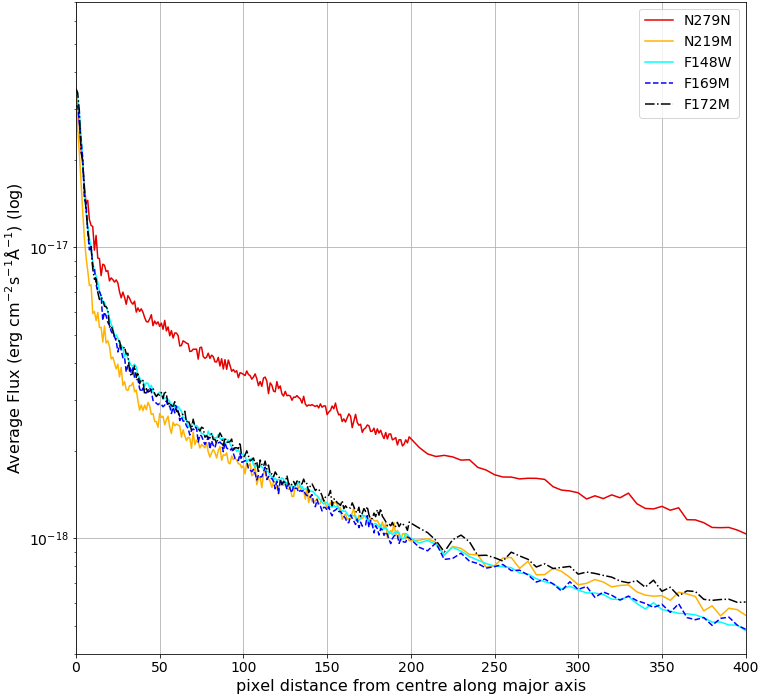}{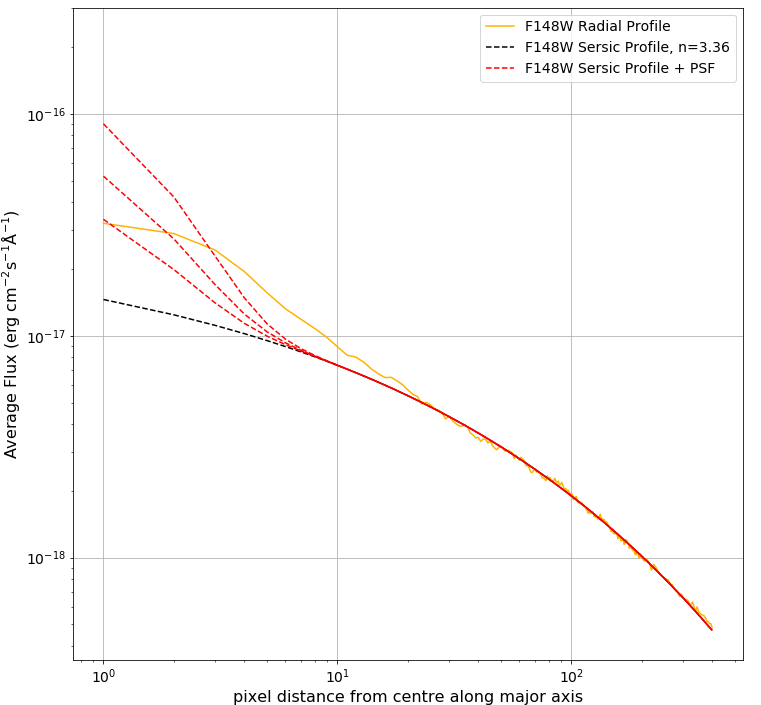}
    \figurenum{4}
    \label{fig:basicellipticalprofiles}
\caption{Left: Elliptical profiles of the bulge (flux per pixel) vs. semi-major axis for the five UVIT filter bands.
The bulge shows a sharp increase in surface brightness toward the centre.
 Right: Sersic fits (black dash line) to the F148W elliptical profile (orange solid line), with added Point Source Functions (PSFs, red dash lines). 
 400 pixels is 167$\arcsec$.}
\end{figure*}

To measure the brightness and UV colours of the bulge we measured average flux per pixel as a 
function of semi-major axis, extracted using elliptical annuli. 
This method was chosen over radial profiles using circular annuli because the images are not circularly symmetric,
and thus the brightness changes significantly with azimuthal angle for circular annuli. 
The brightness profiles using elliptical annuli accounts for the elliptical shape of the bulge projected on the sky.

Pixels (coordinates x,y) were counted within a given ellipse annulus if they satisfied: 
\newline
$2a-0.5< \sqrt{(x-F1x)^2+(y-F1y)^2} + $ 
\newline
$\sqrt{(x-F2x)^2+(y-F2y)^2} \le 2a+0.5$. 
\newline
Here $(F1x,F1y)$ and $(F2x,F2y)$ are the positions of the focii of the given ellipse:
\newline
$F1x = Cx+cos(p)*\sqrt{a^2-b^2}$
\newline
$F1y = Cy+sin(p)*\sqrt{a^2-b^2}$
\newline
$F2x = Cx-cos(p)*\sqrt{a^2-b^2}$
\newline
$F2y = Cy-sin(p)*\sqrt{a^2-b^2}$
\newline
$(Cx,Cy)$ is the $(x,y)$ position of the centre of the bulge on the image, $a$ is the semi-major axis desired, in pixels, 
$b$ the corresponding semi-minor axis, and $p$ the inclination of the major axis from the image's horizontal, in radians. 
The average flux per pixel is calculated by summing the counts/sec in each pixel within a thin elliptical annulus, 
dividing by the number of pixels in the annulus, then using the UVIT calibration to convert count rate per 
pixel to flux per pixel.

To extract flux measurements for the same elliptical annuli of the bulge for all filters,
we used the ellipse parameters vs. semi-major axis from the N279N\_A image analysis. 
As a test, the analysis was repeated using elllipse parameters derived from the F148W\_M image. 
However there was little difference in the resulting elliptical profiles, so the analysis presented here was carried out 
using the N279N\_A ellipse parameters. 
Average flux per pixel was extracted for 240 elliptical annuli, from the centre out to 400 pixels (167$^{\prime\prime}$) 
along the major axis. Examples of these ellipses are drawn in figure \ref{fig:ellipsescontours}. 
One ellipse was used for each of the first 200 pixels along the major axis, and then one ellipse for every 5 pixels for the remaining 200 pixels. 

\begin{table*}[h!]
\begin{center}
\caption{Best-fit bulge profile parameters$^a$.}
\label{tab:chi-squaredtable}
\begin{tabular}{|l|l|l|l|l|l|}
\hline
Single Sersic Fit$^b$ & N279N & N219M & F172M & F169M & F148W\\
\hline
$I0$ & 2.01E-17 & 4.69E-17 & 1.84E-15 &4.28E-16 & 1.84E-16\\
$k$ & 0.278 & 1.26 & 4.15 & 2.79 & 2.01\\
$n$ & 2.54 & 4.74 & 9.07 & 6.79 & 5.55 \\
$\chi^2$ & 104 & 74 & 124 & 151 & 190\\
DOF$^c$ & 236 &  &  &  & \\
\hline
Single Sersic Fit$^d$ & N279N & N219M & F172M & F169M & F148W\\
\hline
$I0$ & 1.70E-17 & 2.17E-17 & 1.34E-16 & 8.05E-17 & 5.16E-17\\
$k$ & 0.213 & 0.736 & 1.90 & 1.45 & 1.07 \\
$n$ & 2.33 & 3.71 &5.69 &4.78 &4.07 \\
$\chi^2$ & 71 & 50 & 80 & 99 & 918 \\
DOF$^c$ & 226 &  &  &  & \\
\hline
Sersic + Gaussian Fit$^b$ & N279N & N219M & F172M & F169M & F148W\\
\hline
$I0$  & 1.67E-17 & 2.13E-17 & 1.29E-16 & 7.17E-17 & 4.86E-17\\
$k$  & 0.206 & 0.723& 1.87 & 1.37 & 1.03\\
$n$  & 2.30 & 3.68 & 5.64 & 4.64 &4.00\\
$Ig$ & 1.37E-17 & 1.9902E-17 & 1.30E-17 &1.10E-17 &1.24E-17\\
$\sigma$ &4.47 &3.97 &4.09 & 4.49 & 4.33\\
$\chi^2$ & 71 & 51 & 81 & 98 & 914\\
DOF$^c$ & 234 &  &  &  & \\
\hline
Double Sersic Fit$^b$ & N279N & N219M & F172M & F169M & F148W\\
\hline
$I0_1$ & 1.62E-17 & 1.46E-17 & 8.29E-17 & 4.81E-17 &3.15E-17\\
$k_{1}$ & 0.195 & 0.509 & 1.53 & 1.10 & 0.762\\
$n_1$ & 2.27 & 3.20 & 5.09 & 4.18 & 3.51\\
$I0_2$ &2.06E-17 & 2.59E-17 & 2.21E-17 & 2.06E-17 & 2.86E-18\\
$k_{2}$ & 0.172 & 0.560  &0.244  & 0.263  & 0.403 \\
$n_2$ & 0.848 &1.39 & 0.948 &1.02 &1.20\\
$\chi^2$ & 69 & 47 & 79 & 94 & 770\\
DOF$^c$ & 233 &  &  &  & \\
\hline
\end{tabular}
\end{center}
\tablecomments{a: The uncertainties in all parameters are $\sim5$\%.}
\tablecomments{b: The fits included data for semi-major axis of 1 to 400 pixels.}
\tablecomments{c:  DOF stands for degrees-of-freedom.}
\tablecomments{d:  The fits included data for semi-major axis of 10 to 400 pixels.}
\end{table*}

\subsection{UV colour-colour diagram}

To study the colour change in the bulge, we produced a colour-colour diagram using the elliptical profiles of flux per
pixel from the above analysis.
The colours chosen were ratio of flux per pixel of F148W to N219M and ratio of N219M to N279N. 
The standard deviation of values of the fluxes from the different pixels in each annulus were used as estimate 
of the flux errors, and propagation of errors was used to calculate the errors in the colours.

\section{Results} \label{sec:results}

\subsection{Elliptical brightness profile analysis}

Analytic functions were fit to the elliptical  surface brightness profiles of the bulge to quantify the shape in a simple manner.
The fits were carried out using $\chi^2$ minimization. 
For these fits the fractional error for each semi-major axis value
was taken as $1/\sqrt N$ with $N$ the number of counts at tht  semi-major axis\footnote{The 
systematic errors in converting counts/s to flux or magnitude (\citealt{2020AJ....159..158T} and  
\citealt{2017AJ....154..128T}) are small compared to the $1/\sqrt(N)$ statistical errors for these narrow (1 pixel wide) elliptical areas.}.

Initially, we fit a Sersic function,  $f(a)$, to the observed surface brightness vs. semi-major axis, $a$. Sersic functions are commonly used to approximate stellar distributions in elliptical galaxies and 
spiral bulges:
\begin{equation}
f(a)=I0\times exp(-k\times a^{1/n})
\end{equation} 
 with $I0$ the central surface brightness, $n$ the Sersic index, and $k$ the e-folding constant
 for decrease of surface brightness.
This is equivalent to the original form of the Sersic function given by \citet{1968adga.book.....S}, 
$f(r)=\Sigma_e \times exp(-b(n)\times[(r/R_e)^{1/n}-1])$ with
$\Sigma_e=I0\times exp(-b(n))$ and $R_e=(b(n)/k)^n$.
  For our data, in units of flux per pixel (erg cm$^{-2}$ s$^{-1}$ A$^{-1}$ pixel$^{-1}$), we use
 units of pixels for $a$.

\begin{figure}[htbp]
    \centering
    \epsscale{1.3}
\plotone{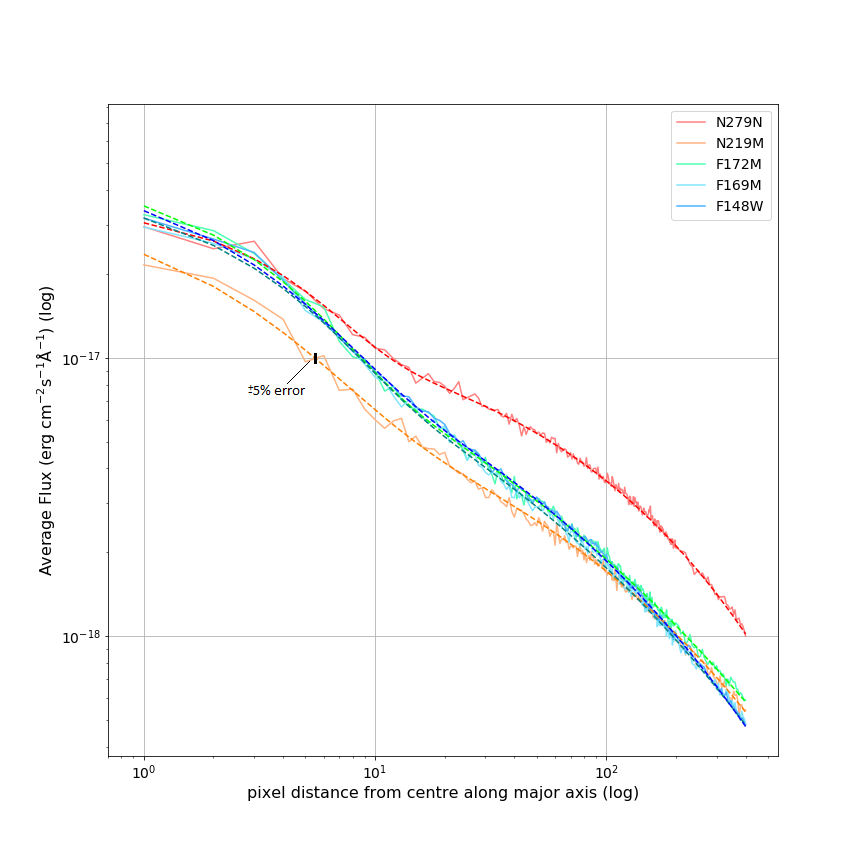}
   \figurenum{5}
\label{fig:GaussFitsBest}
\caption{
  Double Sersic fits (dashed lines) compared to the data (solid lines). The fits include data from 1  pixels to 400 pixels (167$\arcsec$). Data errors range from a minimum 1.3 to 1.8\% for F148W to a 
  maximum of 2.5 to 4.5\% for N279N. Errors for the other filters range between 1.7 to 3.1\%. 
  A $\pm$5\% error bar is shown on the N219M curve for illustration.
  The errors are similar to the differences between the data  points
   and the fit curves.}
\end{figure}

The elliptical profiles generally follow a Sersic shape, with an excess in the center in all filters.
The right panel of Figure~\ref{fig:basicellipticalprofiles} shows the F148W  profile.  
The Sersic function (black dashed line) fits very well the majority of the bulge 
(from semi-major axis of $\simeq 8\arcsec$ to 167$\arcsec$). 
The excess is not consistent with a point source at the centre of the bulge, as shown by the red dashed lines
in  Figure~\ref{fig:basicellipticalprofiles}, which include Sersic plus the radial profile of a point source (the UVIT 
point-spread-function or PSF, \citealt{2017AJ....154..128T}).
The shape of the excess is too broad and flat compared to the PSF.
Higher resolutions are required to study the inner $\sim2\arcsec$, as was done by \citet{2017PhDT.......170L} at optical wavelengths.

Table~\ref{tab:chi-squaredtable} shows the results of single Sersic function fits to the 
bulge, including and omitting the inner 10 pixels ($\sim4\arcsec$) of the bulge.
Comparing the two cases shows that omitting the inner $\sim4\arcsec$ results in a significant reduction in $\chi^2$.
Using the fit with the inner $4\arcsec$ omitted, the Sersic index, n, 
varies with wavelength, with n increasing
from 2.3 to 5.7 for decreasing wavelengths 279 nm $\rightarrow$ 219 nm $\rightarrow$ 172 nm, then decreasing from 5.7 to 4.1 for 172 nm $\rightarrow$ 169 nm $\rightarrow$ 148 nm. 
The fits are acceptable ($\chi^2\lesssim$ degrees-of-freedom or DOF) except for the F148W data, which has the highest signal-to-noise of the five filters.

To better fit the F148W data, and to see if the inner part of the bulge could be fit, we carried out fits 
to the full data (1 to 400 pixels) 
with double-Sersic functions, and Sersic+Gaussian functions. The Gaussian function is:
\begin{equation}
g(a)=Ig\times exp(-a^2/(2\sigma^2))
\end{equation} 
Results are given in Table \ref{tab:chi-squaredtable}. 
For F148W, the Sersic plus Gaussian is significantly better than a single Sersic ($\chi^2$ of 914 vs. 1900),
and the double Sersic better than  Sersic plus Gaussian  ($\chi^2$ of 770 vs. 914).

Figure \ref{fig:GaussFitsBest} shows the double Sersic fits to measured surface brightness for all filters, illustrating that the central bulge
($\lesssim10\arcsec$) and rest of the bulge ($\sim10$ to $167\arcsec$) are well fit. 
The N219M filter observed surface brightness looks different than the other filters: it is faintest
in the inner bulge and approximately matches the shorter wavelength profiles for the outer 
bulge\footnote{The conversion between counts and flux is not the cause for this difference, because
the conversion was well-calibrated (\citealt{2017AJ....154..128T}, \citealt{2020AJ....159..158T}).. 
The intrinsic mean stellar emission spectrum decreases with decreasing wavelength and
 would yield N219M fainter than N279N but brighter than the FUV filters.
However, N219M 
it is at the peak of the UV extinction curve at 220 nm (Fig.3 of \citealt{2007ApJ...663..320F}). At 220 nm
$A_{\lambda}/ A_V$ is larger than for larger or smaller wavelengths by a factor of $\sim$1.6, so  
moderate extinction can make the N219M surface brightness smaller than the FUV surface brightnesses.}

\begin{figure*}[htbp]
\includegraphics[width=10 cm]{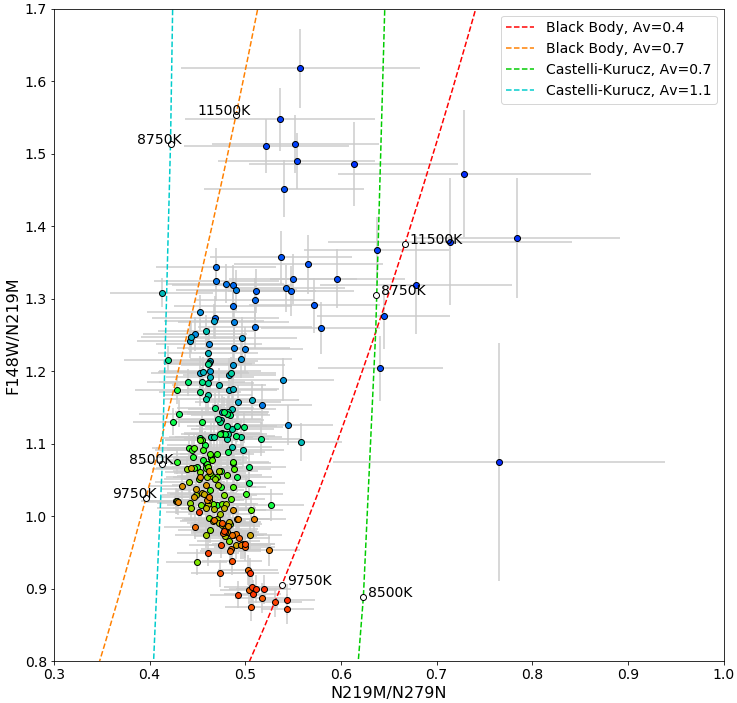}
\includegraphics[width=7.5 cm]{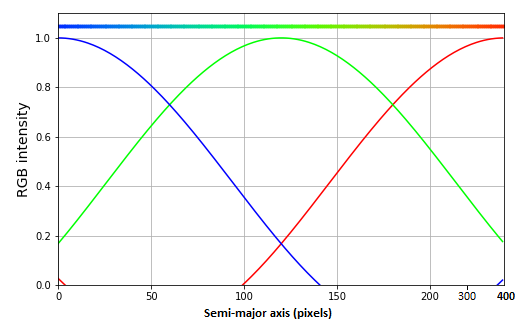}
    \figurenum{6}
\label{fig:c-cdiagram1}
\caption{Left: UV colour-colour diagram of the bulge of M31, for elliptical annuli of different semi-major axes,
 from 0 to 400 pixels (0 to 167$\arcsec$).
  Semi-major axis is encoded by RGB colour, as shown by the panel on the right (the color scale is compressed for
semi-major axes between 200 and 400 pixels because the NUV and FUV colours change slowly for 
semi-major axes $\gtrsim200$.
 The dashed lines show the UV colours for black-bodies and Castelli \& Kurucz stellar atmosphere models of varying 
temperature along the line, with different extinction values for each line.}
\end{figure*} 

\begin{figure*}[htbp]
    \centering
    \epsscale{0.85}
 \plotone{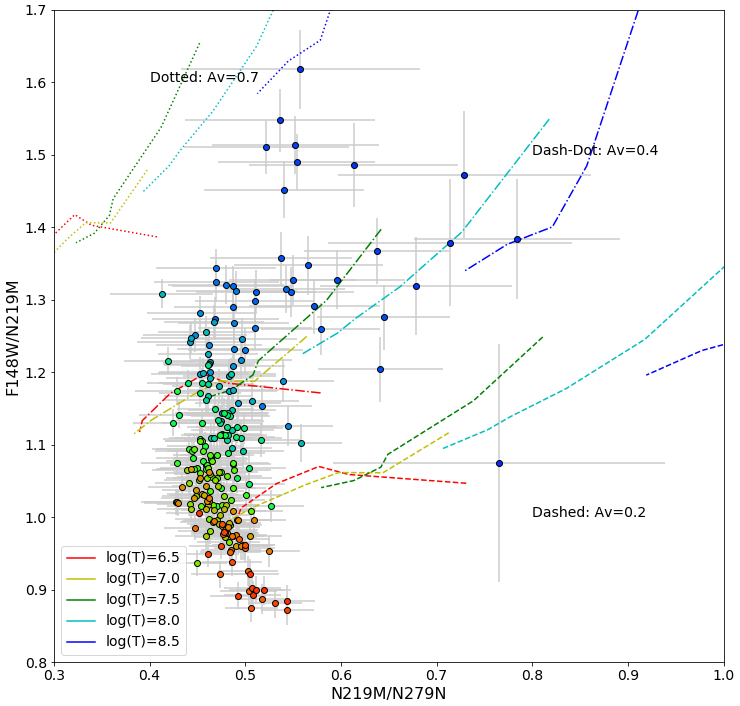}
    \figurenum{7}
\label{fig:c-cdiagram2}
\caption{UV colour-colour diagram of the bulge of M31, for elliptical annuli of different semi-major axis.
The dashed, dash-dot, and dotted lines show the colours calculated from stellar population synthesis models (see
text), with different extinction for each set of 5 lines. 
Stellar metallicity varies along each line, from log(Z/Z$_{\odot}$)=-2.0 to +0.5 (left end to right end), 
and population age, with T in years, is different for each line according to colour.}
\end{figure*}

\subsection{Simple Models for UV colours of the M31 bulge}

\begin{figure*}[htbp]
    \centering
    \epsscale{1.15}
 \plottwo{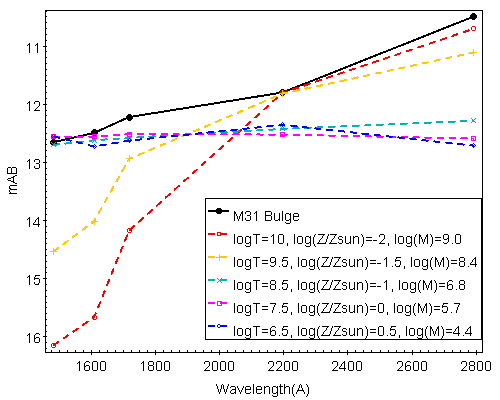}{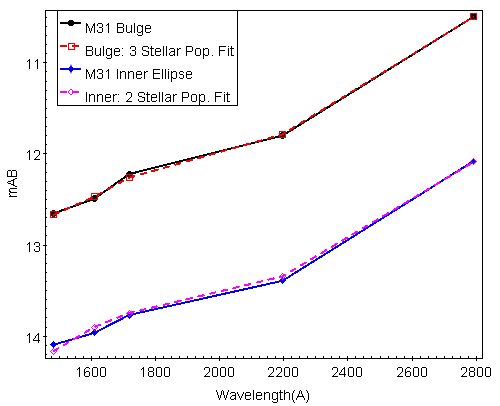}
    \figurenum{8}
\label{fig:1pop}
\caption{Left panel: Single stellar population (SSP) models compared to the data (black line) for the 
Whole  Bulge region (0 to 188$\arcsec$ in semi-major axis, see Table 2) of  M31.
T is the age in years and Z is the metallicity of the population. None of the SSPs can fit the data. Right panel:
data for the Whole Bulge region (black symbols and line) and for the innermost annulus (0 to 58$\arcsec$ in semi-major axis,
blue symbols and line); model of 3 combined SSPs for the Whole Bulge region (red symbols and dashed line); model of 2 combined SSPs for the innermost annulus (magenta symbols and dashed line).}
\end{figure*}

\begin{figure}[htbp]
    \centering
    \epsscale{1.15}
 \plotone{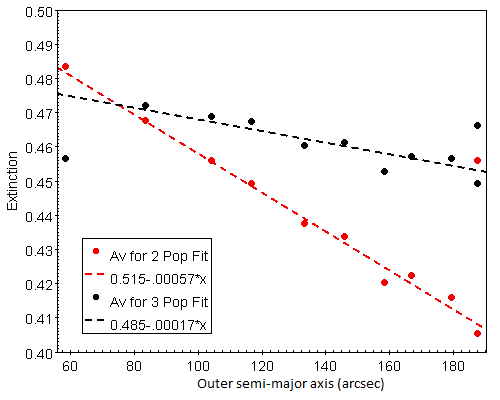}
    \figurenum{9}
\label{fig:Av}
\caption{Best-fit extinction ($A_V$) vs. outer semi-major axis ($a_{out}$) of each annulus  from the 2 SSP fits (red points) and from the 3 SSP fits (black points).  Table 2 gives the inner and outer semi-major axes of each annulus, $a_{in}$  and $a_{out}$. The two points for largest $a_{out}$ are for the outer-most annulus (lower point) and the Whole Bulge region (upper point). Linear functions are fit to $A_V$ vs. $a_{out}$ for all annuli for the 2 SSP model and fit
to all but the innermost annulus for the 3 SSP model.}
\end{figure}

\begin{figure*}[htbp]
    \centering
    \epsscale{1.15}
 \plottwo{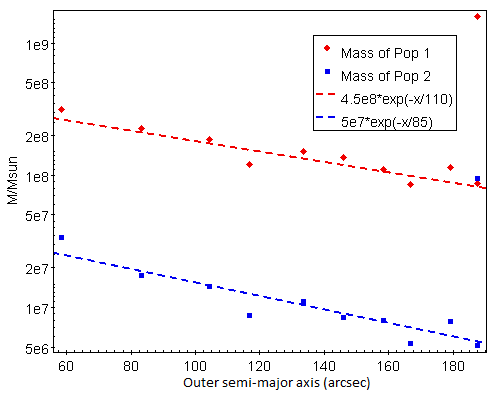}{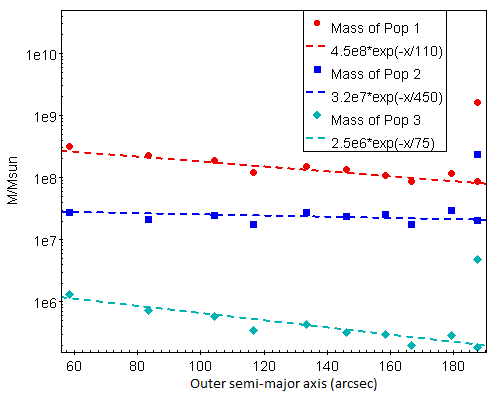}
    \figurenum{10}
\label{fig:masses}
\caption{Left panel: Best-fit SSP masses vs. outer  semi-major axis ($a_{out}$) of each annulus from the 2 SSP fits,
with red points for the older metal-poor population and blue points for the younger metal-rich population. 
  Table 2 gives the inner and outer semi-major axes, $a_{in}$  and $a_{out}$, of the 10 annuli
and of the Whole bulge region. 
Right panel: Best-fit SSP masses vs. outer  semi-major axis ($a_{out}$) of each annulus from the 3 SSP fits,
with red points for the oldest metal-poor population, cyan points for the younger metal-rich population,
and blue points for the intermediate age and metallicity population.  
For each set, there are two points at largest $a_{out}$: the lower mass is for the outermost annulus and the 
higher mass is for the Whole Bulge region. 
Exponential functions are fit to $A_V$ vs. $a_{out}$ for all annuli for the 2 SSP and 3 SSP models.}
\end{figure*}

The UV colour change of the bulge  is shown in Figure~\ref{fig:c-cdiagram1}.
The semi-major axis, $a$, is encoded in a standard RGB colour scheme, with bluest for $a=0$ gradually changing to green at
$a\simeq50\arcsec$, green changing to orange at $a\simeq100\arcsec$ and orange changing to red at the outermost
analyzed ellipse, with $a=167\arcsec$.

The larger errors and higher spread for the innermost ellipses (upper points in the diagram) are caused by
the fewer pixels per annulus for annuli of small semi-major axis.
The colour-colour diagram shows a significant change in the ultraviolet colour of M31's bulge. Points nearest the centre have a greater ratio of F148W flux to N219M flux than those further out, while having a similar ratio of N219M to N279N.
  Thus the FUV-NUV colour (vertical axis) of the bulge becomes "bluer" towards the core.
  Most of the colour change occurs between the centre and half-way out the analyzed region, from semi-major axis of
  0 to $\simeq83\arcsec$.

 Colours for blackbody models and Castelli-Kurucz stellar atmosphere models (\citealt{2003IAUS..210P.A20C}, from the
 STScI website https://www.stsci.edu/ ) are plotted for comparison.
 For the stellar atmospheres, surface gravities were taken as the recommended values for each temperature, 
 and solar metallicity was used. 
 Model colours are calculated as the ratio of filter fluxes determined from a given model spectrum multiplied by the applicable UVIT filter response curve. 
Interstellar extinction is calculated using equations from \citet{2007ApJ...663..320F}
 which include the wavelength dependence for optical and UV. The extinction is parametrized
by $A_V$, given in magnitudes for the V band, and includes internal extinction within the bulge of M31 plus 
foreground extinction in the Milky Way along the line-of-sight toward M31.

The data are consistent with blackbodies between 9,500 and 11,000 K, with extinction values, $A_V$, between 0.4 and 0.7. 
The stellar atmosphere models suggest a lower temperature range, 8500 to 9000 K, and a higher extinction, 
between 0.7 and 1.1.
The stellar models are more realistic if the emission is from stellar atmospheres, particularly if the UV emission 
 primarily comes from a population of unresolved hot stars. 
This change in model temperature can be interpreted as a rise in the average temperature stars towards the centre. 
\citet{2018AJ....156..269L} found evidence for a young hot stellar population in M31's bulge, and this combined with 
other evolved stars are possible causes of the colour changes.

\subsection{Stellar Population Models for UV magnitudes of the M31 bulge}

\begin{table*}[h!]
\begin{center}
\caption{M31 Bulge Stellar Population Fits}
\label{tab:SSPtable}
\begin{tabular}{|l|l|l|l|l|l|l|l|l|}
\hline
\hline
 2 SSP fit$^{a}$\\
\hline
Region & $a_{in}^{b}$ & $a_{out}^{b}$ & $\chi^2$ & M1$^{c}$ & M2$^{c}$ & $A_V$ \\
\hline
Whole & 0 & 187.6 & 4.58 & 1.59E9 & 9.41E7 & 0.456 \\
Ann1  & 0 & 58.4 & 8.77 & 3.13E8 & 3.39E7 & 0.483 \\
Ann2 & 58.4& 83.4 & 7.17 & 2.24E8 & 1.74E7 & 0.467 \\
Ann3 & 83.4 & 104.2 & 5.92 & 1.86E8 & 1.44E7 & 0.456 \\
Ann4 & 104.2 &116.7 & 4.54 & 1.13E8 & 8.63E6 & 0.449 \\
Ann5 &116.7& 133.4 & 4.47 & 1.51E8 & 1.10E7 & 0.438 \\
Ann6 & 133.4& 145.9 & 3.85 & 1.29E8 & 8.32E6 & 0.434 \\
Ann7 & 145.9&158.4 & 4.34 & 1.10E8 & 7.99E6 & 0.420 \\
Ann8 &158.4& 166.7 & 4.03 & 7.14E7 & 5.30E6 & 0.423 \\
Ann9 & 166.7& 179.2 & 5.16 & 1.08E8 & 7.80E6 & 0.416 \\
Ann10 & 179.2& 187.6 & 4.94 & 6.81E7 & 5.14E6 & 0.405 \\
\hline
\hline
3 SSP fit$^{d}$ \\
\hline
Region& $a_{in}^{b}$  & $a_{out}^{b}$ & $\chi^2$ & M1$^{c}$ & M2$^{c}$ & M3$^{c}$ & $A_V$ \\
\hline
Whole  & 0 &  187.6 & 1.01 & 1.59E9 &2.73E8 & 4.76E6 & 0.466 \\
Ann1  & 0 & 58.4 & 1.11 & 3.13E8 &2.73E7  &1.32E6 & 0.457 \\
Ann2 & 58.4 & 83.4  & 1.15 & 2.24E8 & 2.12E7 & 7.17E5 &0.472 \\
Ann3 & 83.4&  104.2  & 1.14 & 1.86E8 & 2.48E7 & 5.84E5 &0.469 \\
Ann4&  104.2  & 116.7 & 0.84 & 1.20E8 &1.77E7 & 3.44E5 &0.467 \\
Ann5 & 116.7& 133.4 & 0.96 & 1.52E8 & 2.73E7 & 4.30E5 &0.460 \\
Ann6 & 133.4 & 145.9 & 0.75 & 1.35E8 & 2.38E7 & 3.18E5 &0.461 \\
Ann7 & 145.9& 158.4 & 0.93 & 1.10E8 & 2.51E7 & 3.03E5 &0.453 \\
Ann8 &158.4& 166.7 & 0.77 & 8.57E7 & 1.77E7 &1.99E5&0.457 \\
Ann9  & 166.7&179.2 & 1.05 & 1.15E8 & 2.95E7 & 2.86E5 &0.457\\
Ann10 & 179.2 &187.6 & 0.82 & 8.59E7 & 2.07E7 & 1.66E5 &0.449 \\
\hline
\end{tabular}
\end{center}
\tablecomments{a: The other best-fit parameters were log(T1)=10, log(Z1/Z$_{\odot}$)=-2,  log(T2)=9, log(Z2/Z$_{\odot}$)=-1.5 for Ann1 to Ann10; for Whole the best-fit parameters were the same except log(Z2/Z$_{\odot}$)=-2.}
\tablecomments{b:  $a_{in}$ and $a_{out}$ are the inner and outer semi-major axes for each annulus.}
\tablecomments{c: Masses are given in units of M$_{\odot}$.}
\tablecomments{d: The other best-fit parameters were log(T1)=10, log(Z1/Z$_{\odot}$)=-2, log(T2)=9.5, log(Z2/Z$_{\odot}$)=-2,   log(T3)=8.5, log(Z3/Z$_{\odot}$)=-0.5, for all regions.}
\end{table*}

To construct more realistic models for the FUV and NUV flux from the M31 bulge, which consists of large numbers
of unresolved stars, it is necessary to consider stellar population models. 
The basic models are simple stellar populations (SSPs) which consist of a set of stars of a single age and metallicity
but with a range of initial masses. 
More complex models can then be constructed by combinations of these SSPs.

Here, we use the SSPs using the Padova stellar models, calculated using the CMD 3.4 online tool at 
http://stev.oapd.inaf.it/cgi-bin/cmd. In CMD 3.4, we used the recommended options: 
PARSEC evolutionary tracks \citep{2012MNRAS.427..127B}, version 1.2S, for pre-main sequence 
to first thermal pulsation or carbon ignition, and COLIBRI \citep{2020MNRAS.498.3283P} for thermal 
pulsation- asymptotic giant branch evolution, ending at total loss of the envelope.
 Circumstellar dust properties were used for M stars and for C stars from \cite{2006A&A...448..181G},
long-period variability along the red giant branch and asymptotic giant branch phases was taken from 
\cite{2019MNRAS.482..929T}, and the initial mass function of \cite{2001MNRAS.322..231K} was used.
For our application, synthetic photometry was calculated in AB magnitudes for the UVIT filter bands.

For comparison to the M31 bulge we calculated models for 8 ages,T, with log(T)= 6.5, 7, 7.5, 8, 8.5, 9, 9.5 and 10.
For each age, we calculated models for metallicities, Z, with log(Z/Z$_{\odot}$)= -2, -1.5, -1, -0.5, 0 and 0.5.
For age log(T)=10 we added log(Z/Z$_{\odot}$)= -2.5 to our grid of models.
The UVIT F148W, F169M, F172M, N219M and N279N AB magnitudes for each star in the population were obtained,
then we combined these to obtain the AB magnitudes for each SSP.

We add extinction, with parameter $A_V$, using the extinction curve of \citet{2007ApJ...663..320F}.
The resulting SSP UVIT colours are compared to the colours of the M31 bulge in  Figure~\ref{fig:c-cdiagram2}.
It is seen that the inner bulge colours (blue points) are similar to SSPs with $A_V\sim$0.4-0.6 and a range of
ages or metallicities. The middle bulge colours (orange points) are similar to SSPs with $A_V\sim$0.2 and 
low age (log(T)$\sim6.5$ to 7 and low metallicities ($\sim-1.5$ to -2).

The above comparison of SSPs with colours is incomplete, 
so we carry out fitting of the M31 bulge FUV-NUV spectral energy distribution (SED) to multiple stellar populations. 
This has the advantages of including data from all five UVIT filters, instead of three, and 
preserves the normalization which is required to obtain the size of the stellar populations.
We extracted the FUV and NUV magnitudes for the whole bulge,
here taken as an outermost ellipse with semi-major axis of 450 pixels (188$\arcsec$).
This large ellipse was divided into 10 annuli of $\simeq$equal-area to examine spatial variations.
The semi-minor axis, b, to semi-major axis, a, ratios (b/a) were taken from the above analysis. 
For the whole and 10 annuli regions, systematic errors in conversion from counts/s to flux and magnitude are
comparable to the $1/\sqrt N$ statistical errors, so were included.

Table~\ref{tab:SSPtable} shows the parameters of the elliptical annuli.
For each area the UVIT magnitudes were fit by minimizing $\chi^2$ with either 1, 2 or 3 SSPs.
The fitting was carried out for each SSP in the grid of log(T) and log(Z/Z$_{\odot}$). For each SSP
the free parameters were mass of the SSP and $A_V$. For the case of 2 or 3 SSP fits, the extinction
was taken to be the same for both (or three) SSPs. 
The best-fit was chosen to be the SSP with the lowest $\chi^2$ of all SSPs in the grid.
For the case of 2 SSPs and 3 SSPs the grids were 4 and 6 dimensional, resp.

None of the single SSP models are able to fit the data. 
The left panel of Figure~\ref{fig:1pop} shows a few of the single SSPs compared to the data for the Whole region
(black line labelled M31 Bulge). The data errors are comparable to the symbol sizes.
The SSP normalization was chosen so that the calculated magnitudes of the SSP did not exceed the data.
It is seen that young SSPs (log(T)$\lesssim$9) are too faint for the longer wavelength data (by $\sim$2 magnitudes) 
and that old SSPs are too faint for the shorter wavelength data.

Including a second SSP improves the fit significantly. 
The right panel of Figure~\ref{fig:1pop} shows the best-fit 2 SSP model (magenta line) for the innermost annulus/ellipse (data shown by blue symbols and solid line).
Adding a third SSP further improves the fit. The Whole region data (top, black symbols and solid line) and the
best-fit 3 SSP model (red symbols and dashed line) are shown. 
The best-fit masses and extinction ($A_V$) for the 2 and 3 SSP models are listed in Table~\ref{tab:SSPtable}.

For the 1 SSP model there are 4 parameters (2 grid parameters and 2 continuous parameters) and 5 data points. 
Best-fit $\chi^2$ values are between 1000 to 2000 for the 10 annuli and whole regions and rule out single SSP models/
For the 2 SSP models there are 7 parameters (4 grid, 3 continuous), more than the number of data points, 
thus a statistically good fit should have $\chi^2\sim1$. 
The $\chi^2$ values are between 4 and 9, indicating a large improvement over the 1 SSP model. 
The 3 SSP models have 10 parameters (6 grid, 4 continuous) and $\chi^2\sim1$ as expected for good fits.
The 3 SSP models show significantly better fits than the 2 SSP models, but they have the disadvantage of
having more parameters than the number of data points in the SED.

\section{Discussion} \label{sec:disc}

M31's bulge, analyzed here out to $\sim170\arcsec$, is elliptical. 
The ellipticity is different in the different NUV and FUV filters (Figure~\ref{fig:GaussFitsAxes}) with
a smaller ellipticity found in the longer wavelength filters (e.g. at semi-major axis of 25$\arcsec$, $a/b\simeq1.17$ 
in NUVN2 vs. 1.27 in F148W). 
 The ellipticity is lowest  near the centre: $a/b=\simeq1.14$: 
 this is apparent on the contour plots (Figure~\ref{fig:ellipsescontours}).
 It increases outward, peaking at $\sim17\arcsec$ major axis in the NUV, and at $\sim25\arcsec$ in the FUV. 
 Further from the centre the ellipticity decreases outward, to a minimum near $\sim55\arcsec$, then increases
outward to  the maximum semi-major axis analyzed here ($\sim95\arcsec$).

Outside of 17$\arcsec$, the ellipticity increases with decreasing wavelength (left panel of Figure~\ref{fig:GaussFitsAxes}). 
Inside of 17$\arcsec$ the ellipticity decreases with decreasing wavelength, i.e. it is approximately reversed, likely 
indicating a different physical component, or different mix of stellar populations, in the central part of the bulge. 
 
The bulge’s inclination changes smoothly with increasing semi-major axis, increasing from $\simeq23^\circ$ reaching a 
peak of $\simeq48^\circ$ in all filters at $\sim25\arcsec$. 
It then slowly decreases outwards to $\simeq33^\circ$, implying a clockwise rotation from centre to semi-major axis of 
17$\arcsec$, followed by a gradual counterclockwise rotation from there outward to the maximum semi-major axis of $95\arcsec$. 
This small amount of rotation (25$^\circ$) is in the contour plots (Figure~\ref{fig:ellipsescontours}) but it is not 
easy to discern. 
The bulge is significantly more circular very near the centre.

The brightness profile of the bulge vs. semi-major axis was extracted using 1 pixel wide elliptical annuli. 
There is a sharp increase of brightness towards the centre.
The profile was fit with a Sersic function, showing an excess at the centre which is more extended
($\simeq2\arcsec$) than a point source (right panel of Figure~\ref{fig:basicellipticalprofiles}).

Thus we fit the bulge with the sum of a Sersic and a Gaussian and the sum of two Sersic functions. 
The double Sersic functions yield the best fit, although the decrease in $\chi^2$ is highly significant only for
the F148W data. 
We conclude that 
the profile is more complex than can be fit with simple analytic functions, such as we have used. 
For no galaxy other than M31 do we have such good photometry in NUV and FUV.
Thus simple analytic fits to a galaxy bulge profile have not tested to this level previously. 

The next step of analysis of the bulge considered spectral changes vs. semi-major axis distance. 
We analysed the colour-colour diagram using flux ratio of F148W to N219M (equivalent to N219M magnitude
minus F148W magnitude) vs. flux ratio of N219M to N279N (equivalent to N279N magnitude
minus N219M magnitude). In this diagram, hotter (brighter at shorter wavelengths) objects are higher in the
diagram, and objects with more extinction are further left. The latter is a result of the peak of the UV extinction
curve at $\simeq$220 nm, matching the N219M filter.

From the colour-colour diagram (Figure~\ref{fig:c-cdiagram1}), the centre of the bulge is seen to be hotter
(blackbody temperature $\simeq$11500 K, or Castelli-Kurucz stellar atmosphere temperature $\simeq$8800 K) 
than the outer parts of the bulge (blackbody temperature $\simeq$9750 K, or Castelli-Kurucz stellar atmosphere temperature $\simeq$8450 K).
The extinction of the bulge is between $A_V=0.4$ and 0.7 for blackbodies, or  between $A_V=0.7$ and 
1.1 for stellar atmospheres. The data errors are such that the outer bulge has distinctly lower extinction than
the middle bulge, but the central bulge has extinction consistent with either middle or outer bulge.

Because the bulge NUV and FUV emission is from a large number (of order $\sim10^9$) of stars,
we compare the colour-colour diagram to the colours calculated for single stellar populations (SSPs).
SSPs with different ages and metallicities are shown. 
Figure~\ref{fig:c-cdiagram2} shows that comparison. 
The apparent trend is that the outer bulge (red points) is more consistent with a young ($\sim10^{6.5-7}$ yr) 
metal poor (log($Z/Z_{\odot})\sim-2$) SSP, the middle bulge (green points) is more consistent with slightly
less young  ($\sim10^{7-7.5}$ yr), less metal poor (log($Z/Z_{\odot})\sim-1$ SSP,
and the inner bulge (blue points)  is more consistent with older ($\sim10^{8.5}$ yr), 
metal poor (log($Z/Z_{\odot})\sim-1.5$) SSP.

We carried out fits to the bulge spectral energy distribution (SED) using all five UVIT filters.
No single SSP is a good fit (Figure~\ref{fig:1pop}). 
This shows that the colour-colour diagram does not contain enough information to properly constrain stellar populations.
A mixture of two SSPs or three SSPs was able to fit the SED of the M31 bulge. 
The three SSP fit is significantly better than the two SSP fit.  

The SEDs were extracted for 10 elliptical annuli in the bulge each with $\simeq$10\% of the whole bulge area, with 
outer semi-major axis of each annulus ranging from $\sim60\arcsec$ to  $\sim190\arcsec$.
For each annulus two SSP and three SSP fits were carried out. The results are shown in
Figure~\ref{fig:Av} for extinction, and Figure~\ref{fig:masses} for masses of the SSPs.
There is a mixture of at least two and probably three different SSPs, based on $\chi^2$ of the fits
in Table~\ref{tab:SSPtable}. This holds for all 10 annuli, i.e., throughout the M31 bulge.

The extinction is variable, decreasing from centre ($<60\arcsec$) to outer part of the bulge (180 to $190\arcsec$)
in a nearly linear manner.
The slope from the two SSP fit is steeper than for the three SSP fit, but the extinction falls in a narrow range 
($A_V=0.4$ to 0.5) for both models. 
The three SSP fits show evidence for a decrease in extinction for the centre. 

The two SSP fit has an old ($10^{10}$ yr) metal poor  (log($Z/Z_{\odot})=-2$)  population plus an intermediate age
($10^{9}$ yr)  less metal poor  (log($Z/Z_{\odot})=-1.5$) population for the whole bulge and for all of the
individual annuli.
The masses of the old and intermediate age populations vary with distance from center but the ratio of intermediate
SSSP mass to old SSP mass is typically $\simeq0.1$.  
Both populations mass vs. distance from centre are well fit by exponential
functions, with scale lengths of 65$\arcsec$ for the intermediate age SSP and 110$\arcsec$ for the old SSP.
The higher concentration of the younger hotter population toward the centre can explain the trend
seen in the colour-colour diagram.

The three SSP fit has an old ($10^{10}$ yr) metal poor  (log($Z/Z_{\odot})=-2$)  population plus an intermediate age
($10^{9.5}$ yr)  metal poor  (log($Z/Z_{\odot})=-2$) population and a young 
($10^{8.5}$ yr)  higher metallicity  (log($Z/Z_{\odot})=-0.5$) population. 
The masses of the three populations vary with distance from center. 
The ratio of intermediate SSP mass to old SSP mass is typically $\simeq0.15$.
The ratio of young SSP mass to old SSP mass is typically $\simeq0.02$. 
The SSP masses vs. distance from centre are well fit by exponential functions, with scale lengths of 
75$\arcsec$ for young SSP, $\sim450\arcsec$ for the intermediate SSP and 110$\arcsec$ for the old SSP.

The dominant old metal poor population has the same scale height and normalization from the two SSP and three SSP fits.
The three SSP fit (with significantly improved $\chi^2$) effectively replaces the $10^9$yr component of the two SSP fit with two separate SSPs of ages $10^{9.5}$yr and  $10^{8.5}$yr. The scale-heights of the latter two populations
are very different: the $10^{9.5}$yr SSP has scale-height 450$\arcsec$ vs. 75$\arcsec$ for the youngest SSP.
The different scale-heights of the 3 SSPs explains the observed colour change of the bulge:
it is hotter near centre where the youngest SSP has the largest fractional contribution. 
 
 \citet{2018MNRAS.475.2754H} carried out hydrodynamical simulations of mergers with M31 and compared the 
 simulations to several sets of observations of M31, including parameters of the Giant Stream, the M31 disc 
 age-dispersion relation, the long-lived 10 kpc ring, the recent and long-lived star formation event, and the 
 slope of the halo profile. They found that a 4:1 major merger where the nuclei merged 2-3 Gyr ago,
 with first passage 7-10 Gyr ago,  could explain many of the observed features of M31. 
 \citet{2018ApJ...868...55M} analyzed data from the Pan-Andromeda Archaeological Survey of M31 to summarize
 the substructures in the stellar halo of M31. Their analysis shows the most distinctive substructures were  produced
 by at least 5 different accretion events in the past 3-4 Gyr. 
 
 The minimum of 5 accretion events \citep{2018ApJ...868...55M} can be
 consistent with merger with a single major galaxy \citep{2018MNRAS.475.2754H} because there were
 a few passages between first passage and  final merger of the nuclei. 
 The passage would have created major streams
 of stars and gas which would cause accretion events at several different times and with different orbital parameters.

 Our SSP fits to the bulge measure star formation in the bulge, and are consistent with the above results. 
 This star formation is caused by accretion of gas or disturbance of ambient gas to induce star formation
 \footnote{This is less likely for the bulge than for the disc.}, thus related to merger and accretion events.
 The $10^{9.5}$yr population comprises 17\%, and the $10^{8.5}$yr population
 comprises 0.3\% of the mass of the old population (Table~\ref{tab:SSPtable}). 
 If there are more than 3 SSPs in the bulge, the current analysis is unable to detect them given the data errors. 
 
 There have been previous studies of the stellar populations in the bulge of M31 
 (\citealt{2018MNRAS.478.5379D},  \citealt{2018A&A...618A.156S} and references therein). 
 Our results are consistent with theirs, in that the bulk of the mass of stars are old (age $\sim$10 Gyr),
 but not consistent in that they find that the old bulge population is metal rich, with [Fe/H]$\sim$0.3 dex. 
 The main difference between the previous studies and this study is that previous studies have used optical observations (optical HST
 photometry, or Lick indices), while we use FUV and NUV band photometry. 
 One possibility is that the current set of SSP models are not accurate enough in the FUV to determine metallicity reliably.
 We plan to update the current study when SSP models which are better verified in the FUV become available.
 
 The spatial structure of the M31 bulge has been studied by \cite{2011ApJ...739...20C}, with primary 
 analysis of the Spitzer IRAC 3.6 micron observations of M31. 
 That study fitted the bulge ignoring the inner few arcseconds containing the nucleus of M31. 
 The fits used a Sersic profile for radial profiles along the major and minor axes or azimuthally averaged radially profiles. 
 They used two fitting methods: non-linear least squares and Monte Carlo Markov Chain. 
 The results (their Table 2) were a Sersic index, n, between 1.7 and 2.4 and an effective radius, 
 $R_e$, between 0.5 and 1.1 kpc.
 Both n and $R_e$ had typical errors of 10\% .
 In comparison, they fit the 2-D image using the GALFIT software \citep{2010AJ....139.2097P} 
 and obtained similar results: n between 1.9 and 2.0 and $R_e$ between 0.9 and 1.0 kpc.
 Our longest wavelength is 279 nm, far shorter than the 3.6 micron data analyzed by \cite{2011ApJ...739...20C}. For 279 nm we find n values of 2.33, 2.30 and 2.27 for single Sersic, Sersic plus Gaussian or double Sersic fits, respectively (second, third and fourth sections of Table 1).
 These values are only slightly higher than those from  \cite{2011ApJ...739...20C}.
 For the shorter NUV 219 nm and the three FUV bands, we find significantly higher Sersic 
 indices (3.2 to 5.6). 
 The fact that the Sersic indices are quite different for NUV and FUV should not be too surprising, 
 given that the shorter wavelengths are highly sensitive to the rare young and hot stars.
 These we have shown are more concentrated in the centre of the bulge (most clearly shown here in Fig.~\ref{fig:masses}). 
 
\section{Conclusions}  \label{sec:conc}

We have carried out observation of the bulge of the Andromeda Galaxy in NUV and FUV wavelengths using 
the UVIT instrument on the AstroSat Observatory.
M31's bulge is elliptical in shape with small but distinctive changes in ellipticity and position angle with distance
from the nucleus of M31. 

The bulge is centrally concentrated in UV, and approximately follows a Sersic profile.
The nucleus (in the innermost $4\arcsec$) is complex, 
as known from optical studies \citep{2017PhDT.......170L}.
For the brightness profiles in all filters but F148W, 
the bulge well fit  by Sersic plus Gaussian or double Sersic functions.
For F148W, with its high signal-to-noise, none of the attempted functions gave statistically good fits,
but the double Sersic had the best-fit, and gave no systematic residuals vs. radius, which may indicate that the
bulge has small scale structure or clumps which are not fit by any smooth function. 

The UV colour-colour diagram shows that the bulge becomes significantly bluer towards the centre. 
We carried out an analysis of the FUV -NUV SEDs using SSPs for the whole bulge and the bulge subdivided
into 10 elliptical annuli. Our best fit model has three SSPs at each radius and yields the same best-fit
ages and metallicities vs. radius for these SSPs. 
There is a dominant old SSP ($10^{10}$yr, log($Z/Z_{\odot})=-2$), an intermediate-mass, intermediate-age 
SSP ($10^{9.5}$yr, log($Z/Z_{\odot})=-2$) and a young low-mass SSP ($10^{8.5}$yr, log($Z/Z_{\odot})=-0.5$).
The different SSPs have exponential mass profiles vs. radius with different scale-heights for each SSP.
The ages and masses of the SSPs are consistent with an active merger history for M31, as known from studies
of M31 using optical data.

The FUV and NUV data from UVIT/AstroSat are useful to measure the young stellar populations which 
emit at these wavelengths. Future studies of regions beyond the bulge promise to turn up more important properties 
of the ages and spatial distributions of stellar populations in M31.  

{\large \bfseries \noindent Acknowledgements}
\newline This project is undertaken with the financial support of the Canadian Space Agency and of the Natural Sciences and Engineering Research Council of Canada. This publication uses data from the AstroSat mission of the Indian Space Research Institute (ISRO), archived at the Indian Space Science Data Center (ISSDC).

\end{document}